# Geofence and Network Proximity

Dmitry Namiot  
Lomonosov Moscow State University  
Faculty of Computational Math and Cybernetics  
Moscow, Russia  
e-mail: dnamiot@gmail.com

Manfred Sneps-Sneppe  
Ventspils University College  
Ventspils International Radio Astronomy Centre  
Ventspils, Latvia  
e-mail: manfreds.sneps@gmail.com

*Abstract*— **Many of modern location-based services are often based on an area or place as opposed to an accurate determination of the precise location. Geo-fencing approach is based on the observation that users move from one place to another and then stay at that place for a while. These places can be, for example, commercial properties, homes, office centers and so on. As per geo-fencing approach they could be described (defined) as some geographic areas bounded by polygons. It assumes users simply move from fence to fence and stay inside fences for a while. In this article we replace geo-based boundaries with network proximity rules. This new approach let us effectively deploy location based services indoor and provide a significant energy saving for mobile devices comparing with the traditional methods.**

## I. Introduction

Geo-fencing enables remote monitoring of geographic areas surrounded by a virtual fence (geo-fence), and automatic detections when tracked mobile objects enter or exit these areas [1]. A huge set of LBS (location based services) use geo-fence observation as a key feature. Location plays a basic role in context-aware applications. Geo-fences are user-defined areas defined around a Location. Locations here are cities, towns, other identifiable landmarks as well as vehicle parks of the user organization. Usually, the user is able to define the bounding of geo-fence area. For example, in simplest case it is just a radius defines some circular area. On practice, in the vehicle tracking system, a vehicle is determined to be at a particular Location if it is within this geo-fence (e.g., within the given radius for circular area).

Any geo-fence implementation requires obviously some form of location monitoring. Technically, this monitoring could be performed either right on the mobile device or via some centralized scheme (e.g., telecom operator observes the location for own subscribers).

The main sources for user's raw coordinates on mobile phones as Global Positioning System (GPS) and Wireless Positioning System (WPS) using cell tower and Wi-Fi access points (AP) [2].

One of the biggest and well known problems with the location monitoring is energy consumption. It is, probably, the biggest limitation factor. Typical battery capacity of smart phones today is barely above 1000 mAh (e.g., the lithium-ion battery of HTC Dream smart phones has the capacity of 1150 mAh). GPS, the core enabler of LBS, is power-intensive, and its aggressive usage can cause complete drain of the battery within a few hours [3]. A typical GPS invocation consists of a locking period and a sensing/reporting period. The lengths of these two periods are about 4-5 seconds and 10-12 seconds, respectively. More importantly, the average power draws for the two above-mentioned periods are about 400 mW and 600 mW, respectively. For a typical battery capacity of 1000 mAh such high power consumption is very expensive as continuous GPS sensing can deplete the battery in merely 6 hours [4].

Figure 1 illustrates battery depletion test with GPS mode on

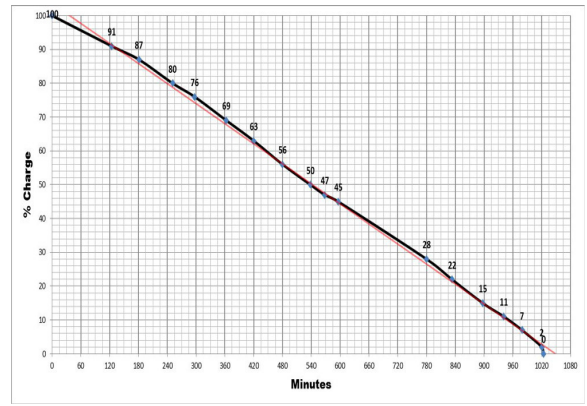

Figure 1.    Battery depletion [5]

Figure 2 compares GPS and non-GPS modes as well as illustrates the power spikes:

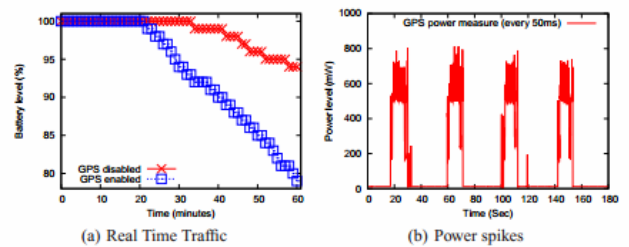

Figure 2.    GPS vs. non-GPS mode [4]

The battery drain effect could be presented more dramatically in case of several applications that work in parallel. Note, that several independent location monitoring applications present the more realistic picture.

Many research papers declare the goal to develop frameworks that continuously provides location context with minimum energy consumption.

We can mention here, for example, deploying fingerprints to recognize semantic places with high level accuracy using radio beacons (e.g., cell towers, WiFi APs, and Bluetooth), counting the surrounding factors (e.g., light, texture, and sound patterns) – so called context. As per classical definition [6], context-related information can consist of a users profiles and preferences, their current location, the type of connection that to the mobile network, the type of wireless device being used, the objects that are currently in the user's proximity, and/or information about their behavioral history. Actually, most of the authors define context awareness as complementary element to location awareness, whereas location may serve as a determinant for resident processes. By this reason, all the context-aware applications are linked to location exchange.

To optimize energy consumption for continuous sensing, various approaches have been proposed. These include sensor selection by movement detector using accelerometers [7, 8], minimizing energy consumption within accuracy requirements [9, 10], utilizing a prediction-based approach [11], etc.

We can select the following common directions (areas) for energy saving during the location monitoring:

1) Adaptive selection of location sensing mechanisms. Actually, it should be selected dynamically (e.g., GPS or network fingerprints). Location sensing mechanisms could have performance tradeoffs in terms of accuracy, power consumption, and availability.

2) Usage of context information. Modern LBS should be context-aware too.

3) Adding cooperation for multiple LBS on client's side. They should communicate by some way in order to avoid redundant location sensing invocations [12].

But in general, any client side monitoring is and always will be energy consuming operation. The more prospect area by our opinion is the centralized location monitoring [13]. It is one of the few areas where telecom operators can get advantages over Internet companies and effectively use own base of connected devices. One possible example: Sprint Geofence API [14]. Another example is Open API platform for Alcatel-Lucent [15]. Unfortunately, this prospect line in location monitoring is not elaborated yet from the practical point of view. The biggest problem by our opinion is the lack of common standards. One possible candidate for such standard in telecom was Parlay, but at this moment we cannot name one widely accepted candidate. That is why most of the scientific papers and practical implementations are devoted to the client side location monitoring.

The rest of the paper is organized as follows. Section II contains an analysis of existing projects devoted to network proximity. In Section III, we consider our Spotique service and related applications.

## II. NETWORK PROXIMITY

The main idea behind our Spotique service described below is the replacement of geo data with network proximity. We will try to describe geo fence (geographically restricted areas) with network proximity rules and replace geo locations monitoring with networks proximity monitoring. The reasons behind this movement are very transparent.

At the first hand, it will work indoor. At the second, it is based on the actions, performed by the most smart-phones anyway. Most of the mobile users keep Wi-Fi on all the time. And Wi-Fi scanning is a part of the network proximity. So, from the energy saving point of view, there are no extra operations.

Network scanning is centralized. So, for any mobile phone all the installed LBS based on the network proximity will use the same data (share the same processes) automatically (see above-mentioned remark about cooperation for multiple LBS on the client side).

Network proximity based systems support dynamic LBS. It is described in [16], for example. If our "location" is linked by some way to Wi-Fi access point (network node), than not only this node could move. It could be opened (closed) dynamically right on the mobile phone:

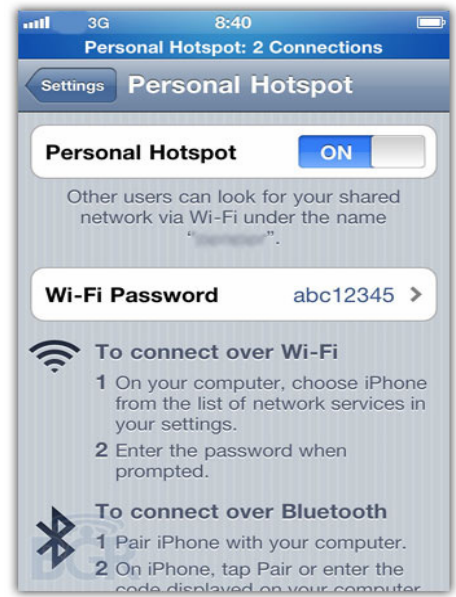

Figure 3. Wi-Fi hot spot right on the mobile

One example for LBS based on the network proximity principles is SpotEx [17]. In this concept, any existing or even specially created wireless network node could be used as a presence sensor, which can open (discover) access to some dynamic or user-generated content. As a key service point, SpotEx introduces an external database with some rules (productions or if-then operators) related to the Wi-Fi access points. Typical examples of conditions in our rules are: AP (access point) with SSID *Café* is visible for mobile device; RSSI (signal strength) is within the given interval, etc. Based on such conclusions, we then deliver (make visible) user-defined messages to mobile terminals. In other words, the visibility of the content depends on the network context (Wi-Fi network environment).

Technically, SpotEx presents proximity information via some set of rules. Each rule is a logical production (if-then operator). The conditional part includes the following objects:

Wi-Fi network (SSID, mac-address)
RSSI (signal strength - optionally)
Time of the day (optionally)
ID for the client (mac-address)

It means that collection of all rules is a set of operators like:

IF *IS_VISIBLE*('mycafe') AND *FIRST_VISIT()* THEN {present the coupon info } [18]

LifeTag [19] uses collected database of so called Wi-Fi "fingerprints", including MAC addresses and the received signal strengths (RSSI) of nearby access points for discovering the user's behavioral patterns.

What could be used as fingerprint? One simplest approach could be based on the time any particular Wi-Fi access point is visible from the mobile phone [20]. The MAC addresses of visible access point let us logically estimate the location ("not far from that access point"). The mobile application on the phone can record periodically MAC addresses from received Wi-Fi beacons. A fingerprint is acquired by computing the fraction of times each unique MAC address was seen over all recordings. A tuple of fractions (each tuple element corresponding to a distinct MAC address) forms the Wi-Fi fingerprint of that place.

Fingerprint matching is performed by computing a metric of similarity between a test fingerprint and all candidate fingerprints. The comparison between two fingerprints, $f_1$ and $f_2$, is performed as follows. Denote M as the union of MAC addresses in $f_1$ and $f_2$. For a MAC address m ∈ M, let $f_1(m)$ and $f_2(m)$ be the fractions computed as above. Then the similarity S of $f_1$ and $f_2$ is computed as:

MinMax(m) = min($f_1(m),f_2(m)$)/max($f_1(m), f_2(m)$)

$$S = \sum_{m \in M} (f_1(m)+f_2(m)) * \text{MinMax}(m)$$

The intuition behind this metric is to add a large value to S when a MAC address occurs frequently in both $f_1$ and $f_2$.

The purpose of the fraction is to prevent adding a large value if a MAC address occurs frequently in one fingerprint, but not in the other. Note, that this calculation does not use signal strength measurement at all.

For geo-fence analogue we can compare current fingerprint and pre-recorded fingerprints for boarding points. Any given metric for similarity let us describe proximity (e.g., "close enough").

A classical approach to Wi-Fi fingerprinting [21] involves RSSI (signal strength). The basic principles are transparent. At a given point, a mobile application may hear ("see") different access points with certain signal strengths. This set of access points and their associated signal strengths represents a label ("fingerprint") that is unique to that position. The metric that could be used for comparing various fingerprints is k-nearest-neighbor(s) in signal space. It means that two compared fingerprints should have the same set of visible access points and they could be compared by calculating the Euclidian distance for signal strengths.

Fingerprinting is based on the assumption that the Wi-Fi devices used for training and positioning measure signal strengths in the same way. Actually, it is not so (due to differences caused by manufacturing variations, antennas, orientation, batteries, etc.). To account for this, we can use a variation of fingerprinting called ranking. Instead of comparing absolute signal strengths, this method compares lists of access points sorted by signal strength. For example, if the positioning scan discovered ($SS_A$; $SS_B$; $SS_C$) = (-20; -90; -40), then we replace this set of signal strengths by their relative ranking, that is, ($R_A$; $R_B$; $R_C$) = (1; 3; 2) [21]. As the next step, we can compare the relative rankings by using the Spearman rank-order correlation coefficient [22].

We can use signal strength features for distance estimation in terms of the Euclidean distance in signal strength space and the Tanimoto coefficient [23].

As a prerequisite we compute the vector of average signal strength per access point $S'_x$ from the list of signal strength vectors $S_x$. In the Euclidean version the distances are defined as follows for each pair of average signal strength vectors $S'_a$, $S'_b$, with entries for non-measurable access points in either vector set to -100dBm:

$d_{a,b} = \|S'_a - S'_b\|$

For the Tanimoto coefficient version, the distance is computed as follows so the value increases as the vectors becomes more dissimilar:

$d_{a,b} = 1 - (S'_a \cdot S'_b)/(\|S'_a\|^2 + \|S'_b\|^2 - S'_a \cdot S'_b)$

Technically, it means that we can describe our geo-area as a set of basic point with statically calculated fingerprints. And later we can compare current fingerprint for mobile device with our basic fingerprints. By the similar schemes work almost all Wi-Fi based positioning systems. But there are two main problems. At the first hand, the task for creating basic "Wi-Fi mail stones" could be expensive. Also we will need to recalculate them every time our network environment is changed. At the second, we will face the same problems with energy consumption during the client side calculations. To overcome this we can use a fact that for the proximity calculation we do not need the distance. For the simple proximity calculation we can use some form of graph for signal strength versus distance for one Wi-Fi access point

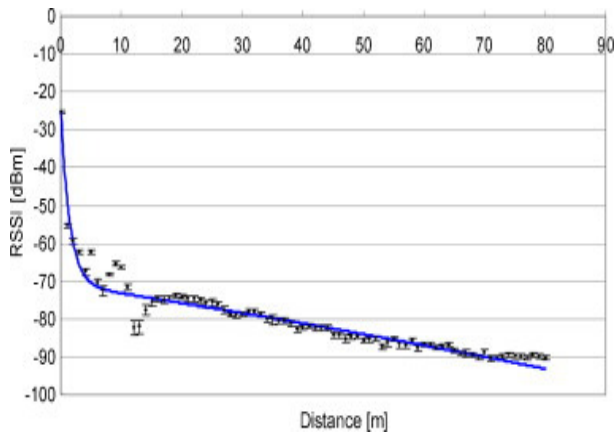

Figure 4.  RSSI vs. distance [24]

And for several Wi-Fi access points we can combine individual metrics.

As per energy consuming, we think that the most proper direction is to remove measurements processing from the mobile phone completely. It is what our Spotique service is about.

### III. SPOTIQUE SERVICE

Our Spotique service let broadcast hyper-local message to mobile clients. A typical usage scenario is: a user with a Wi-Fi device walking near a shop sees an ad for the hot offer on his Wi-Fi device, and also captures a coupon from the shop. The user then enters the shop and redeems his coupon by displaying it on the screen. By the same principles we can distribute information is Smart City projects, etc.

Spotique disconnects location-related calculations from mobile phone and uses server-side proximity detection based on Wi-Fi beacons. Wi-Fi client (mobile phone in our case) can periodically send so called probe request frame [25]. As per Wi-Fi spec, a station sends a probe request frame when it needs to obtain information from another station. For example, a client would send a probe request to determine which access points are within range. It is so called passive Wi-Fi tracking.

One benefit of the beacon-based approach is that it is implicitly location-aware. We are dealing with devices whose locations are known and whose accessibility is limited by the propagation of 802.11 signals. This approach works regardless of whether the client Wi-Fi devices are already connected to an existing Wi-Fi network, or the client is completely disconnected from all Wi-Fi networks. For mobile phones it is completely enough just to keep Wi-Fi networks mode switched on.

An additional benefit of this approach is that we eliminate the need to explicitly locate the client, which as a side-effect improves the privacy model. This scheme does not require from clients to send messages that explicitly reveal their location.

Technically, such external beacon-based monitoring can provide the following information about Wi-Fi based devices in proximity:

MAC – address
RSSI (signal strength)

There are several out-of-the-shelf components that can provide probe request detection [26, 27]:

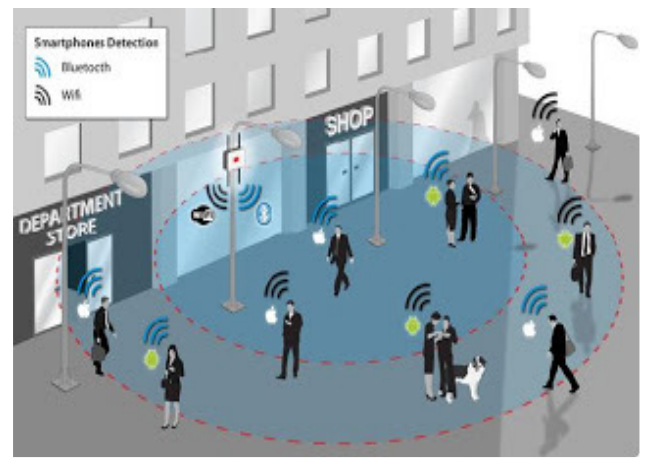

Figure 5.  Wi-Fi beacons detection

In our scheme we will use RSSI info for getting in-proximity devices and MAC-address for the identification. From the database on detected Wi-Fi devices we can get MAC-addresses for units in the proximity. Note, that for the security reasons we can replace the real MAC-address with some hash.

But how can we detect our subscribers among them? For doing this we will deploy Cloud Messaging [28]. Google Cloud Messaging for Android (GCM) is a service that lets developers send data from servers to their applications on Android devices. This could be a lightweight message telling the Android application that there is new data to be fetched from the server (for instance, a movie uploaded by a friend), or it could be a message containing up to 4kb of payload data (so apps like instant messaging can consume the message directly). The GCM service handles all aspects of queuing of messages and delivery to the target Android application running on the target device. There is the similar service in Apple's world too (Apple Push Notification).

GCM allows 3rd-party application servers to send messages to their Android applications. GCM deployment depends on two things: Application ID and Registration ID. Application ID assigned to the Android application that is registering to receive messages. The Android application is identified by the package name from the manifest. This ensures that the messages are targeted to the correct Android application.

Registration ID is unique ID issued by the GCM servers to the Android application that allows it to receive messages. Once the Android application has the registration ID, it sends it to the 3rd-party application server, which uses it to identify each device that has registered to receive messages for a given Android application. In other words, a registration ID is tied to a particular Android application running on a particular device.

An Android application on an Android device doesn't need to be running to receive messages. The system will wake up the Android application via Intent broadcast when the message arrives, as long as the application is set up with the proper broadcast receiver and permissions.

And here is the main idea for Spotique: during the registration for GCM collect MAC-address for Wi-Fi tracking. Spotique presents mobile application that lets users subscribe to the topics (read – local businesses) they are interested. Local businesses define our topics. Each subscription tights in the server side database the following things: topic, MAC-address for the subscriber (for his mobile phone) and Registration ID for GCM.

Each time when passive tracking system at the local business location obtained MAC-address for in-proximity mobile device, we can check that MAC-address against subscription database. And as soon as the user (mobile phone, actually) is subscribed, we can push to his phone our custom message, using his Registration ID from the same database.

In this schema we eliminate the need to explicitly locate the client. This scheme does not require from clients to send (or publish in some social network) messages that explicitly reveal their location.

Note again, that for improving the privacy we do not need even to save in our database original MAC-addresses. It is enough to keep some hash-code instead of the real address.

For Apple Push Notification (APN) service each device establishes an accredited and encrypted IP connection with the service and receives notifications over this persistent connection. If a notification for an application arrives when that application is not running, the device alerts the user that the application has data waiting for it. Developers ("providers") originate the notifications in their server software. The provider connects with APNs through a persistent and secure channel while monitoring incoming data intended for their client applications. When new data for an application arrives, the provider prepares and sends a notification through the channel to APNs, which pushes the notification to the target device. For the future development with APN we will keep the same principles for subscription as the above described GCM model.

IV. CONCLUSION

This paper discusses geo-fence limitations for mobile applications and offers as a replacement network proximity model. We discuss several models related to Wi-Fi proximity. Namely, they are fingerprints and rule based proximity. We describe a new location based service Spotique. It is based on passive Wi-Fi tracking and Cloud Messaging. This new approach let us effectively deploy location based services indoor and provide a significant energy saving for mobile devices comparing with the traditional methods. In Spotique we eliminate the need to explicitly locate the client, which as a side-effect improves the privacy model. This scheme does not require from clients to send messages that explicitly reveal their location. The proposed approach eliminates one of the biggest concerns for location based systems adoption – privacy.

REFERENCES

[1] F. Reclus and K. Drouard Geofencing for fleet & freight management Intelligent Transport Systems Telecommunications,(ITST),2009 9th International Conference on pp. 353-356

[2] A.LaMarca et al. Place lab: Device positioning using radio beacons in the wild. In Lecture Notes in Computer Science, volume 3468, pp. 116–133, Munich, 2005.

[3] I. Constandache, S. Gaonkar, M. Sayler, R. R. Choudhury, and L. Cox. Enloc: Energy-efficient localization for mobile phones. In Proceedings of IEEE INFOCOM Mini Conference '09, Rio de Janeiro, Brazil, 2009.

[4] Z. Zhuang, K.-H. Kim, and J. Singh Improving Energy Efficiency of Location Sensing on Smartphones MobiSys'10, June 15–18, 2010, San Francisco, California, USA, Copyright 2010 ACM 978-1-60558-985-5/10/06

[5] http://forums.watchuseek.com/f296/garmin-fenix-ongoing-review-several-parts-746366-11.html Retrieved: Nov, 2012

[6] N. Hristova and G. M. P. O`Hare, "Ad-me: Wireless Advertising Adapted to the User Location, Device and Emotions," in Thirty-Seventh Hawaii International Conference on System Sciences (HICSS-37), 2004.

[7] H. Lu et al. The jigsaw continuous sensing engine for mobile phone applications. In Proc. 8th ACM Conf. Embedded netw. Sens. Syst.,SenSys'10, pp. 71–84. ACM, 2010.

[8] J. Paek, J. Kim, and R. Govindan. Energy-efficient rate-adaptive gpsbased positioning for smartphones. In Proc. 8th MobiSys'10, pp. 299–314. ACM, 2010.

[9] D. Namiot "Geo messages", Ultra Modern Telecommunications and Control Systems and Workshops (ICUMT), 2010 International Congress pp. 14-19 DOI: 10.1109/ICUMT.2010.5676665

[10] B. Priyantha, D. Lymberopoulos, and J. Liu. LittleRock: Enabling Energy-Efficient Continuous Sensing on Mobile Phones. IEEE Pervas. Comput., 10(2), pp. 12–15, Apr-Jun 2011.

[11] L. Song, D. Kotz, R. Jain, and X. He. Evaluating next-cell predictors with extensive wi-fi mobility data. IEEE Trans. Mobile Comput., 5(12), pp. 1633 –1649, 2006.

[12] Y. Chon, E. Talipov, H.Shin, and H.Cha Mobility Prediction-based Smartphone Energy Optimization for Everyday Location Monitoring SenSys'11, November 1–4, 2011, Seattle, WA, USA, Copyright 2011 ACM 978-1-4503-0718-5/11/11

[13] Schneps-Schneppe, M., & Namiot, D. Telco Enabled Social Networking: Russian Experience. In BALTIC CONFERENCE (p. 33).

[14] Sprint Geofence API: http://developer.sprint.com/dynamicContent/geofence/ Retrived: Nov, 2012

[15] Open API platform http://www2.alcatel-lucent.com/application_enablement/ Retrieved: Nov, 2012

[16] Y. Daradkeh, D Namiot, and M. Sneps-Sneppe Spot Expert as Context-Aware Browsing, Journal of Wireless Networking and Communications, vol.2, N. 3, 2012, pp. 23-28

[17] D. Namiot and M. Schneps-Schneppe, "About location-aware mobile messages", NGMAST, 2011, pp. 48-53 DOI: 10.1109/NGMAST.2011.19

[18] D. Namiot and M. Sneps-Sneppe Proximity as a Service, Future Internet Communications (BCFIC), 2012 2nd Baltic Congress on, pp. 199-205 DOI: 10.1109/BCFIC.2012.6217947

[19] J. Rekimoto, T. Miyaki, and T. Ishizawa. LifeTag: WiFi-based Continuous Location Logging for Life Pattern Analysis. in LOCA. 2007

[20] M. Azizyan, I. Constandache, and R.Roy, "SurroundSense: mobile phone localization via ambience fingerprinting", MobiCom '09 Proceedings of the 15th annual international conference on Mobile computing and networking, pp. 261-272, DOI: 10.1145/1614320.1614350

[21] Y. Chen, Y. Chawathe, A. LaMarca, and J. Krumm. "Accuracy characterization for metropolitan-scale Wi-Fi localization", In ACM MobiSys, 2005.

[22] A.Stuart The correlation between variate-values and ranks in samples from a continous distribution. British Journal of Statistical Psychology Vol. 7, Issue 1, pp. 37–44

[23] M. Kjaergaard, M. Wirz, D. Roggen, and G.Troster Mobile sensing of pedestrian flocks in indoor environments using WiFi signals Pervasive Computing and Communications (PerCom), 2012 IEEE International Conference on pp. 95 – 102

[24] W. Janga and W. Healyb Wireless sensor network performance metrics for building applications, Energy and Buildings Vol. 42, Issue 6, pp. 862–868, DOI: 10.1016/j.enbuild.2009.12.008

[25] R. Chandra, J. Padhye, L. Ravindranath, and A.Wolman A Beacon-Stuffing: Wi-Fi without Associations Mobile Computing Systems and Applications, 2007. HotMobile 2007. Eighth IEEE Workshop on, pp.53-57

[26] Smartphones movement track http://www.technologyreview.com/view/427687/if-you-have-a-smart-phone-anyone-can-now-track/ Retrieved: Nov, 2012

[27] Meshlium Xtreme http://www.libelium.com/products/meshlium Retrived: Nov 2012

[28] J.Hansen, T.Gronli, G.Ghinea Cloud to Device Push Messaging on Android: A Case Study, Advanced Information Networking and Applications Workshops (WAINA), 2012 26th International Conference on, 2012 pp. 1298 – 1303